\documentclass[12pt]{article}
\usepackage{a4}
\usepackage{amsfonts}
\usepackage[dvips]{graphicx}
\begin{document}
\begin{titlepage}
\vskip0.5cm
\begin{flushright}
HUB-EP-00/39\\
\end{flushright}
\vskip0.5cm
\begin{center}
{\Large\bf Eliminating leading corrections to scaling in the}
\vskip 0.3cm
{\Large\bf 3-dimensional $O(N)$-symmetric $\phi^4$ model:}
\vskip 0.3cm
{\Large\bf $N=3$ and $4$}
\end{center}
\vskip 1.3cm
\centerline{
Martin Hasenbusch
}
\vskip 0.4cm
\centerline{\sl  Humboldt-Universit\"at zu Berlin, Institut f\"ur Physik,}
\centerline{\sl  Invalidenstr. 110, D-10115 Berlin, Germany}
\vskip 0.3cm
\centerline{e-mail: hasenbus@physik.hu-berlin.de}

\vskip 0.4cm
\begin{abstract}
We study corrections to scaling in the $O(3)$- and $O(4)$-symmetric
$\phi^4$ model 
on the three-dimensional simple cubic lattice with nearest neighbour
interactions. For this purpose, we use Monte Carlo
 simulations in connection with a finite size scaling method.
We find that there exists a finite value of the coupling $\lambda^*$,
for both values of $N$, where leading corrections to scaling vanish.
As a first application, we compute the critical exponents
$\nu=0.710(2)$ and $\eta=0.0380(10)$ for $N=3$ and $\nu=0.749(2)$ 
and $\eta=0.0365(10)$ for $N=4$.
\end{abstract}
\end{titlepage}
\section{Introduction}
At a second order phase transition various quantities diverge with power laws.
E.g. the magnetic susceptibility behaves as 
\begin{equation}
\label{power}
\chi \sim C_{\pm} \; |t|^{-\gamma} \;\;\;,
\end{equation}
where $t=(T-T_c)/T_c$ is the reduced temperature. The subscripts $+$ and $-$
indicate the high and low temperature phase, respectively. 
$\gamma$ is the critical exponent of the magnetic susceptibility.

The universality hypothesis says that for all systems within
a given universality
class the exponent $\gamma$, as other critical exponents, takes exactly the 
same value. 
Note that the amplitudes $C_{\pm}$ depend on the details of the system, 
while the ratio $C_+/C_-$ is universal; i.e. it takes the same value for all 
systems within a universality class.
A universality class is characterized by the spatial
dimension of the system, the range of the interaction and the symmetry of the
order parameter. See ref. \cite{WiKo}.

A central goal of the study of critical phenomena is 
to obtain precise estimates of 
universal quantities like the critical exponents or universal amplitude 
ratios.  

Often so called lattice spin models are used to study  critical 
phenomena. The Ising model is the proto-type of such models.
In three dimensions, where no exact solution of these models is available, 
the most precise results are obtained by the analysis of 
high temperature series or Monte Carlo 
simulations. For universal quantities, a similar accuracy can be obtained 
with field-theoretic methods like the $\epsilon$-expansion or perturbation theory
in three dimensions.  For a detailed discussion see text-books on 
critical phenomena; e.g. refs. \cite{cardy,Domb}.

The accuracy of estimates of universal quantities extracted from high 
temperature series expansions or Monte Carlo simulations is limited by 
corrections to the power law (\ref{power}):
\begin{equation}
\label{correction}
 \chi = C_{\pm} \; |t|^{-\gamma} \;\;
 \left(1 \;+\; a_{\pm} \; |t|^{\theta} \;+\; ... \right) \;\;\;,
\end{equation}
where the correction exponent $\theta=\nu \omega \approx 0.5$ is universal.
While the exponent $\theta$ is universal, the amplitudes of corrections
depend on the parameters of the system. Hence it is a very natural idea 
to search for parameters at which (non-analytic) corrections to scaling vanish. 

In the present work, 
we study  the $O(N)$-invariant $\phi^4$ model on simple cubic lattices with 
periodic boundary conditions. The lattices have the linear extension $L$ in 
all three directions. The classical Hamiltonian is given by
\begin{equation}
\label{action}
H=  -  \beta \sum_{<xy>}  \vec{\phi}_x \cdot \vec{\phi}_y \; 
  \; + \sum_x \; \vec{\phi_x}^2 \; +\; \lambda
       \sum_x \; (\vec{\phi}_x^2-1)^2 \;\;\;,
\end{equation}
where $\vec{\phi}_x \in \mathbb{R}^N$ and $<xy>$ 
denotes a pair of nearest neighbour sites
on the lattice. We study the canonical ensemble; 
the partition function is given by
\begin{equation}
 Z = \int \mbox{D} [\phi] \;\; \exp(-H) \;\;\;,
\end{equation}
where $\int \mbox{D} [\phi]$ is a short hand for the $N \times L^3$ dimensional
integral over all components of the field at all lattice points.

In the limit $\lambda = \infty$, the so called $O(N)$-invariant non-linear 
$\sigma$ model is recovered. In this limit the last term of the Hamiltonian forces
the field to unit-length: $\vec{\phi}_x^2 = 1$. 
$\lambda=0$ gives the exactly solvable
Gaussian model. Along a critical line $\beta_c(\lambda)$, the model undergoes 
a second order phase transition. For all $\lambda>0$, at given $N$, these 
transitions belong to the same universality class. Hence the critical exponents,
including the correction exponent $\theta$, are the same for all values of 
$\lambda>0$. However, in eq.~(\ref{correction}) the correction amplitudes $a_{\pm}$
become functions of $\lambda$.
Hence, there is a chance to find a zero:
$a_{\pm}(\lambda^*)=0$.
Since leading corrections to scaling are due to a unique irrelevant scaling 
field $u_3$, it follows that $\lambda^*$ is unique for all quantities.

Already in refs. \cite{ChFiNi,NiRe} it was suggested that for models that
interpolate between the Gaussian and the Ising model, there exists one
value of the interpolation parameter, where leading corrections to scaling
vanish. These models were studied with high temperature series expansions.

More recently, this idea has been picked up in connection with Monte Carlo
simulations and finite size scaling. In refs. \cite{spain,HaPiVi} a finite 
size scaling method has been proposed that allows to compute $\lambda^*$ in 
a systematic way. While the details of the implementation of ref. \cite{spain} 
and \cite{HaPiVi} are different, the underlying physical idea is the same.

This programme has been successfully implemented for $N=1$ (Ising universality
class) \cite{spain,HaPiVi,Ha99} and $N=2$ ($XY$ universality class)
 \cite{HaTo99,XYneu}.
As a result, for these universality classes, the most precise estimates
for critical exponents from Monte Carlo
simulations were obtained.

In a turn, the results for $\lambda^*$ obtained from Monte Carlo have been 
used as input for the analysis of
high temperature series of these models. In ref. \cite{pisa} the 
Ising universality class and in refs. \cite{XYneu,pisaXY} the $XY$
universality class
was studied. As a result, the accuracy of the critical exponents was 
further improved and in addition, a large set of universal 
amplitude combinations
has been obtained with unprecedented precision.

In the present work we like to answer the question whether  this programme
can be extended to larger values of $N$. 
In ref. \cite{pisa} it was argued that for $N>3$ there 
exists no such $\lambda^*$ for the Hamiltonian eq.~(\ref{action}) on 
simple cubic lattices.
The argument 
is based on the large $N$ expansion and the analysis of high temperature 
series expansions \cite{milano}.

The paper is organized as follows. In section 2 we briefly review the 
finite size scaling method that has been applied to eliminate corrections 
to scaling.
The Monte Carlo algorithm is discussed in section 3. Details of the 
simulations are given in section 4.
In section 5 the numerical results
are presented. Next we compare our results for the critical exponents 
with those given in the literature. Finally we give our conclusions.

\section{The finite size scaling method}
\subsection{Phenomenological couplings}
Here, phenomenological couplings  are used
to locate  $\lambda^*$. 
Nightingale, in his seminal paper on the phenomenological renormalization
group \cite{Nightingale}, 
proposed to use such a quantity to locate $\beta_c$ and to compute 
the renormalization group exponents.
Phenomenological couplings
are invariant under  renormalization group transformations.  Hence, they 
are well suited to detect corrections to scaling.
The proto-type of such a quantity is the so called Binder-cumulant 
\cite{binder}:
\begin{equation}
 U_4 \;=\; \frac{<(\vec{m}^2)^2>}{<\vec{m}^2>^2} \;\;\;,
\end{equation}
where
\begin{equation}
 \vec{m} \; =\; \frac{1}{V} \; \sum_x \; \vec{\phi}_x
\end{equation}
is the magnetisation of the system. Note that also higher moments of the 
magnetisation could be considered.

Frequently 
the second moment correlation length divided by the linear extension
of the lattice $\xi_{2nd}/L$ has been studied.
The second moment correlation length is defined 
by
\begin{equation}
\xi_{2nd} \;=\; \sqrt{\frac{\chi/F-1}{4 \; \sin(\pi/L)^2}} \;\;\;,
\end{equation}
where
\begin{equation}
\chi \;=\; \frac{1}{V} \;
 \left\langle \left(\sum_x \; \vec{\phi}_x \right)^2 \right\rangle
\end{equation}
is the magnetic susceptibility and
\begin{equation}
F \;= \; \frac{1}{V} \;   \left \langle
|\sum_x \exp\left(i \frac{2 \pi x_1}{L} \right) \vec{\phi}_x |^2 \;\;
\right \rangle
\end{equation}
is the Fourier-transform of the two-point correlation function at
the lowest non-vanishing momentum. In order to reduce the statistical
error, we averaged the results of all three directions of the lattice.
Note that Nightingale \cite{Nightingale} studied $\xi/L$,  
where $\xi$ is the exponential correlation length of a system of the 
size $L^{D-1} \times \infty$.

The third quantity that we study is the ratio $Z_a/Z_p$
of the partition function $Z_a$ with anti-periodic boundary
conditions in one of the three directions and $Z_p$ with periodic
boundary conditions in all directions. 
Anti-periodic boundary conditions mean that the term
$\sum_{<xy>} \vec{\phi}_x \cdot \vec{\phi}_y$ in the Hamiltonian
is multiplied by $-1$
for $x=(L_1,x_2,x_3)$ and $y=(1,x_2,x_3)$.
This ratio can be measured with the
help of the boundary-flip algorithm, which is a version of the cluster algorithm.
The boundary-flip algorithm was introduced in ref. \cite{Ha93} for the Ising model.
In \cite{GH-94} the authors have generalized this method
to the case of $O(N)$-invariant non-linear $\sigma$ models. As in ref. \cite{Ha99}
we use a version of the algorithm that only measures $Z_a/Z_p$ and does not 
perform the flip to anti-periodic boundary conditions.
For a recent discussion of the algorithm see ref. \cite{XYneu}.

\subsection{The finite size method}
Below we shall briefly recall the theoretical basis of the finite size scaling
method that has been developed in refs. \cite{spain,HaPiVi,Ha99}. 
See also ref. \cite{XYneu}.

Let us denote the phenomenological couplings by $R$.  We define a quantity $\bar{R}$
based on a pair of phenomenological couplings $R_1$ and $R_2$.

First we define $\beta_f$ as the value of $\beta$, where, 
at given $\lambda$ and $L$, $R_1$ takes the fixed value $R_{1,f}$:
\begin{equation}
\label{betafeq}
 R_1(L,\lambda,\beta_f) = R_{1,f} \;\;.
\end{equation}
Hence $\beta_f$ is a function of $L$ and $\lambda$. In the language of 
high energy physics, this is our ``renormalization condition''.

Next we define 
\begin{equation}
\label{rbardev}
\bar{R}(L,\lambda) \equiv R_2(L,\lambda,\beta_f) \;\;.
\end{equation}
The behaviour of this quantity gives us direct access to corrections to scaling,
as we shall see below. In the following we will frequently refer 
to $\bar{R}(L,\lambda)$ as ``$R_2$ at $R_{1,f}$''.

Let us consider linearized renormalization group 
transformations in the neighbourhood of 
the fixed point.
In the following, we take explicitly 
into account the thermal scaling field $u_t$ and the leading irrelevant
scaling field $u_3$. In addition, we consider $u_4$ as representative 
of sub-leading irrelevant scaling fields. For the definition of scaling fields
see e.g. the text-book \cite{cardy}. 

In oder to derive the finite size scaling behaviour of $R$ we block 
the lattice to a fixed size $L'$. 
The 
scaling fields $u_t'$, $u_3'$ and $u_4'$ of the blocked system are given by 
\begin{equation}
\label{ut}
u_t'  \;=\; a(\lambda,\beta) \; L^{1/\nu} \;\;\;
\end{equation}
with $a(\lambda,\beta_c) =0 $ by definition. The irrelevant scaling fields 
behave as 
\begin{equation}
\label{u3}
u_3'  \;=\; u_{3}^{(0)}(\lambda) \; L^{-\omega} 
\;+\; O(\beta-\beta_c(\lambda)) \;\;
\end{equation}
and
\begin{equation}
\label{u4}
u_4'  \;=\; u_{4}^{(0)}(\lambda) \; L^{-\omega_2} 
\;+\; O(\beta-\beta_c(\lambda)) \;\;\;.
\end{equation}
Here we have absorbed the constant factors $L'^{-1/\nu}$, $L'^{\;\omega}$
and $L'^{\;\omega_2}$ into $a(\lambda,\beta)$, $u_{3}^{(0)}(\lambda)$ and 
$u_{4}^{(0)}(\lambda)$, respectively. 
We always simulate at a vanishing external field. Hence 
also the corresponding scaling field $u_h$ vanishes.
While the value of $\omega\approx 0.8$ is well established 
(see e.g. \cite{GUZI}), the knowledge of $\omega_2$ is rather limited. 
Using the ``scaling field method", which is a version of Wilsons ``exact
renormalization group", the authors of ref. \cite{NR-84} find  
$\omega_2 = 1.78(11)$ for $N=3$. Following the figure given in ref. \cite{NR-84}
the value of $\omega_2$ for $N=4$ is similar to that of $N=3$.
In addition, there are corrections due 
to the breaking of the rotational invariance by the lattice. These 
corrections have an $\omega_{rot} \approx 2$ \cite{CPRV-98}. In the case of the 
Binder-cumulant and the second moment correlation length we have to 
expect corrections due to the analytic background of the magnetic 
susceptibility. This amounts to corrections with an 
exponent $\omega_{back} = 2 - \eta \approx 2$.
In our numerical analysis, we will always pessimistically assume 
$\omega_2 = 1.6$ when 
we estimate systematic errors due to sub-leading corrections.
Since $R$ keeps its value under renormalization group transformations 
we get
\begin{equation}
 R(\beta,\lambda,L) \;=\; R'(u_t',u_3',u_4',...)
\end{equation}
Note that we are allowed to skip $L'$ from the list of arguments of $R'$ since 
$L'$ is fixed.
Expanding $R'$ in the scaling fields 
at the fixed point $\{u_j'=0\}$
gives
\begin{equation}
\label{rexpand}
 R \; = \; R^* \; 
   +D_t \; u_t' \;\; 
   +D_3 \; u_3' \;\;
   +D_4 \; u_4' \;\; 
   + \; ... \;
\end{equation}
with
\begin{equation}
D_i \;\equiv \; \left. \frac{\partial R'}{\partial u_i'} 
\right|_{\{u_j'=0\}} \;\;\;. 
\end{equation}
Inserting eqs.~(\ref{ut},\ref{u3},\ref{u4}) into  eq.~(\ref{rexpand}) yields
\begin{equation}
 R \; = \; R^* \;
 + D_t \; a(\lambda,\beta_f) \; L^{1/\nu} \; 
 + \; c(\lambda) \; L^{-\omega} \;
 + \; d(\lambda)\; L^{-\omega_2} \;+\; ... \;\;\;, 
\end{equation}
where $c(\lambda)=D_3 u_3^{(0)}(\lambda)$ and 
$d(\lambda)=D_4 u_4^{(0)}(\lambda)$.
Hence 
\begin{equation}
a(\lambda,\beta_f) \; L^{1/\nu} \;=\;  D_{1,t}^{-1} \; 
\left(R_{1,f} \; - \; R_1^* \;  - \; c_1(\lambda) \; L^{-\omega}
- \; d_1(\lambda) \; L^{-\omega_2}
\right) \;+ \;... \;\;\; .
\end{equation}
The subscript $1$ has been introduced to indicate that the quantities 
are related to $R_1$.
Taking $R_2$ at $\beta_f$ gives
\begin{equation}
\label{mastereq}
 \bar{R}(L,\lambda) \; = \; \bar{R}^* \; + \bar{c}(\lambda) \; L^{-\omega} \;
 + \; \bar{d}(\lambda) \; L^{-\omega_2}
 \; + \; ...
\end{equation}
with 
\begin{equation}
\bar{R}^* \; =\; R_2^* \;+\; \frac{D_{2,t}}{D_{1,t}} \;(R_{1,f}-R_1^*) 
\;\;\;,
\end{equation}
\begin{equation}
\bar{c}(\lambda) \;=\; c_2(\lambda) \;-\; 
\frac{D_{2,t}}{D_{1,t}} \;  c_1(\lambda) \;\;\;
\end{equation}
and $\bar{d}$ is given by an analogous expression. 

In eq.~(\ref{mastereq}) we have neglected corrections that are quadratic 
in the scaling fields. The most important of these is 
$b \; \bar{c}(\lambda)^2 \; L^{-2 \omega}$. The exponent $2 \omega$ has roughly
the same value as $\omega_2$. However, since $\bar{c}(\lambda)$ appears 
quadratically, we can safely ignore this term in the neighbourhood of 
$\lambda^*$.

Eliminating leading corrections to scaling means in terms of eq.~(\ref{mastereq})
to find the zero of $\bar{c}(\lambda)$.
One can imagine various numerical implementations to find this zero.
Here we followed the strategy used in refs. \cite{Ha99,HaTo99}. We have simulated
the models close to the critical line for several values of $\lambda$ for various
lattices sizes. The results are then fitted by an ansatz motivated by 
eq.~(\ref{mastereq}). The function $\bar{c}(\lambda)$ is then approximated by 
interpolation between the $\lambda$-values, where simulations have been 
performed.

In the following,
 we will always use either $Z_a/Z_p$ or $\xi_{2nd}/L$ as $R_1$ and 
$U_4$ as $R_2$. Note that in ref. \cite{Ha99} we have only used $R_1=Z_a/Z_p$
and in ref. \cite{HaTo99} only $R_1=\xi_{2nd}/L$. Using both quantities
gives us 
better control over systematic errors introduced by sub-leading corrections.

\section{The Monte Carlo Algorithm}
The $O(N)$-invariant 
non-linear $\sigma$ models were simulated with a cluster algorithm.
For finite $\lambda$, following Brower and Tamayo \cite{BrTa}, 
additional updates with a local Metropolis-like algorithm were
performed to allow fluctuations of the modulus of the field $\vec{\phi}_x$.
Below, we give the details of the  cluster algorithm and the local update.

\subsection{The wall-cluster update}
Several variants of the cluster algorithm have been proposed in the 
literature. The best know are the (original) Swendsen-Wang algorithm 
\cite{SW} and the single-cluster algorithm of Wolff \cite{Wolff}.
Here we have used the wall-cluster algorithm of ref. \cite{HaPiVi}.
In ref. \cite{HaPiVi}, for the 3D Ising model, a small gain in
performance compared with the single-cluster algorithm was found.
Such a comparison for $N>1$ remains to be done.
The main reason for using the wall-cluster algorithm here is that the 
wall-cluster update can be combined with the measurement of the ratio 
of partition functions $Z_a/Z_p$. 
In the wall-cluster update all clusters are flipped that intersect a
plane of the lattice.  

The definition of a cluster is the same 
as in the  Swendsen-Wang  or the single-cluster algorithm.
Following Wolff \cite{Wolff},  
only the sign of one component $\phi_x^{(p)}$ of the field is changed  
in a single step of the algorithm.
This leads to the probability
\begin{equation}
\label{pdelete}
p_d = \mbox{min}[1,\exp(-2 \beta \phi_x^{(p)} \phi_y^{(p)})]
\end{equation}
to delete the link $<xy>$. The links that are not deleted are frozen.
A cluster is a set of sites that is connected by frozen links.

In ref. \cite{Wolff} the component of the field parallel to a randomly 
chosen direction is used. Here we choose the $1^{st}$, $2^{nd}$, ... or 
$N^{th}$ component of $\vec{\phi}$.
This simplifies the implementation of the cluster update
and the measurement of $Z_a/Z_p$. Also CPU-time is saved since in one 
update step only one component of the field has to be accessed.  
In order to compensate the algorithmic disadvantage of this restricted 
choice, we perform a global rotation of the field after a certain number of 
cluster updates.

\subsection{The local update of the $\phi^4$ model}
We sweep through the lattice with a local Metropolis-type updating scheme.

A proposal for a new field at the site $x$ is  generated by
\begin{equation}
\phi_x'^{(i)} = \phi_x^{(i)} \; + \; c \; (r^{(i)}-0.5) \;\;\;,
\end{equation}
where the $r^{(i)}$  are random numbers that are uniformly distributed
in $[0,1)$, $i$ runs from 1 to $N$.
The proposal is accepted with the probability
\begin{equation}
A = \mbox{min}[1,\; \exp(-H'+H)] \;\;\;,
\end{equation}
where $H'$ is the Hamiltonian of the proposed field $\phi'$ and 
$H$ the Hamiltonian of the original field.
The step-size $c$ is adjusted such that the acceptance rate is about $1/2$.
After this Metropolis step, we  perform at the same site an
over-relaxation step:
\begin{equation}
\label{overrelax}
\vec{\phi}_x' \;=\; \vec{\phi}_x \;
- \;2 \; \frac{(\vec{\phi}_x \cdot \vec{\phi}_n) 
\;\vec{\phi}_n}{\vec{\phi}_n^2} \;\;\;.
\end{equation}
Note that this step takes very little CPU-time. Hence it is likely that
its benefit out-balances the CPU-cost. For lack of time, 
we did not carefully check this point.

\subsection{The update cycle}
\label{cycle}
Finally let us summarize the complete update cycle:

\begin{itemize}

\item local update sweep \\
(This step is omitted for the non-linear $\sigma$ models)
\item
global rotation of the field 
\item
$3 \times N$  wall-cluster updates
\end{itemize}
The sequence of the $3 \times N$ wall-cluster updates is given by
the wall in 1-2, 1-3 and 2-3 plane. In each of the three cases,
an update is performed for all $N$ components of the field. For each of the 
three orientations of the wall we performed a measurement of $Z_a/Z_p$.

\section{The simulations}
First we have simulated the $O(3)$-invariant non-linear $\sigma$ model at 
the best estimate of $\beta_c=0.693002(12)$ of ref. \cite{Ballesteros}.
We used lattices of size $L=6$, $8$, $12$, $16$, $24$, $32$ and $48$.
We have performed $25 \times 10^6$ measurements for $L=6$ up to $L=32$ and 
$10^7$ measurements for $L=48$.

The $N=3$ $\phi^4$ model was simulated at $\lambda=2.0$, $4.5$ and $5.0$
on lattices of the linear size $L=6$, $8$, $12$, $16$, $24$ and $32$. For
$\lambda=4.5$ we simulated in addition $L=48$.  In all cases we performed 
$10^7$ measurements.

We have simulated the $O(4)$-invariant non-linear $\sigma$ model at
the estimate of $\beta_c=0.935861(8)$ of ref. \cite{Ballesteros}.
We studied lattices of size $L=6$, $8$, $12$, $16$, $24$, $32$ and $48$.
We have performed $25 \times 10^6$ measurements for $L=6$ up to $L=16$,
$2 \times 10^7$ measurements for $L=24$,
$14 \times 10^6$ measurements for $L=32$,
and $105 \times 10^5$ measurements for $L=48$.

The $N=4$ $\phi^4$ model was simulated at $\lambda=8.0$, $12.0$ and 
$14.0$ on lattices of the linear size $L=6$, $8$, $12$, $16$, $24$ and $32$.
For $\lambda=12.0$ we simulated in addition $L=48$.  As for $N=3$, 
we performed $10^7$ measurements for each parameter set.

A measurement was performed after one update cycle. (See section \ref{cycle}.)

In all cases listed above, we performed $10^5$ update 
cycles before measuring.

In order to reduce the amount of data that is written to disc, we 
averaged during the simulation the results of 5000 measurements. These 
averages were saved.

In the case of the $\phi^4$ model we have 
simulated at estimates of $\beta_f$ from $Z_a/Z_{p,f}$, which were obtained 
from smaller lattice sizes that have been simulated before and/or  
short test-simulations.

In our analysis the observables are needed as a function of $\beta$. 
Given the large statistics we did not use the reweighting technique.
Instead we used the Taylor-expansion up to the third order. The 
coefficients were obtained from the simulation. We always checked 
that the $\beta_f$ are sufficiently close to the $\beta$ of the 
simulation such that the error from the truncation of the Taylor-series
is well below the statistical error.

As random number generator we have used our own implementation of 
G05CAF of the NAG-library. The G05CAF is a linear congruential random 
number generator with modulus $m=2^{59}$, multiplier $a=13^{13}$ and 
increment $c=0$.

As a check of the correctness of the program and the quality of the random 
number generator we have implemented the following two non-trivial relations 
among observables:
\begin{equation}
\label{phicheck}
0 = \frac{1}{2} \beta \sum_{y.nn.x} \langle \vec{\phi}_x \vec{\phi}_{y} \rangle
   \;-\;   \langle \vec{\phi}_x^2  \rangle
   \;-\; 2 \lambda \;  \langle (\vec{\phi_x}^2 -1) \; \vec{\phi}_x^2 \rangle
   \;+\; \frac{N}{2}
\end{equation}
and
\begin{equation}
\label{rotate}
0 \;=\;
\beta\;   \left\langle \left( \sum_{y.nn.x}
[\phi_x^{1} \phi_y^{2}  - \phi_x^{2} \phi_y^{1}]
\right)^2 \right \rangle
- \; \frac{2}{N} \; \sum_{y.nn.x} \langle \vec{\phi}_x \vec{\phi}_y
\rangle \;\;\;,
\end{equation}
where $y.nn.x$ indicates that the sum runs over the six nearest neighbours of $x$.
In order to enhance the statistics, we have summed  
eqs.~(\ref{phicheck},\ref{rotate}) over all sites $x$.
We found that for all our simulations these equations are satisfied within the 
expected statistical errors.

Most of the simulations were performed on 450 MHz Pentium III PCs. 
For our largest lattice size $L=48$, one update cycle 
plus a measurement takes 0.67 s, 0.58 s, 0.86 s and 0.75 s for the 
$N=3$ $\phi^4$ model at $\lambda=4.5$, $\beta=0.68622$, the 
$O(3)$-invariant non-linear $\sigma$ model at $\beta=0.693002$,
the $N=4$ $\phi^4$ model at $\lambda=12.0$, $\beta=0.90843$ and the
$O(4)$-invariant non-linear $\sigma$ model at  $\beta=0.935861$, 
respectively.

In total, the whole study took  about two years 
on a single 450 MHz Pentium III CPU.

\section{Analysis of the data}
\subsection{Corrections to scaling}

In a first step of the analysis we have estimated
$\xi_{2nd}/L^*$ and $Z_a/Z_p^*$ by fitting the $O(3)$- and $O(4)$-invariant
non-linear 
$\sigma$ model data with the ansatz
\begin{equation}
\label{RSansatz}
R(\beta_c) \;=\;R^* \; +\; c \; L^{-\omega} \;\;\;,
\end{equation}
where $\beta_c$, $R^*$ and $c$ are the  parameters of the fit.
We have fixed $\omega=0.8$.
As result we obtain $Z_a/Z_{p}^* \approx 0.196$ and $\xi_{2nd}/L \approx 
0.564$ for the $O(3)$ model and 
$Z_a/Z_p^* \approx 0.1195$ and $\xi_{2nd}/L^* \approx 0.547$
for the $O(4)$ model. In the following we shall use these numbers to set 
$R_{1,f}$. I.e. 
$Z_a/Z_{p,f}=0.196$ and $\xi_{2nd}/L_f=0.564$ for $N=3$ and 
$Z_a/Z_{p,f}=0.1195$ and $\xi_{2nd}/L_f=0.547$ for $N=4$.

\begin{figure}[tb]
\begin{center}
\includegraphics[width=12cm]{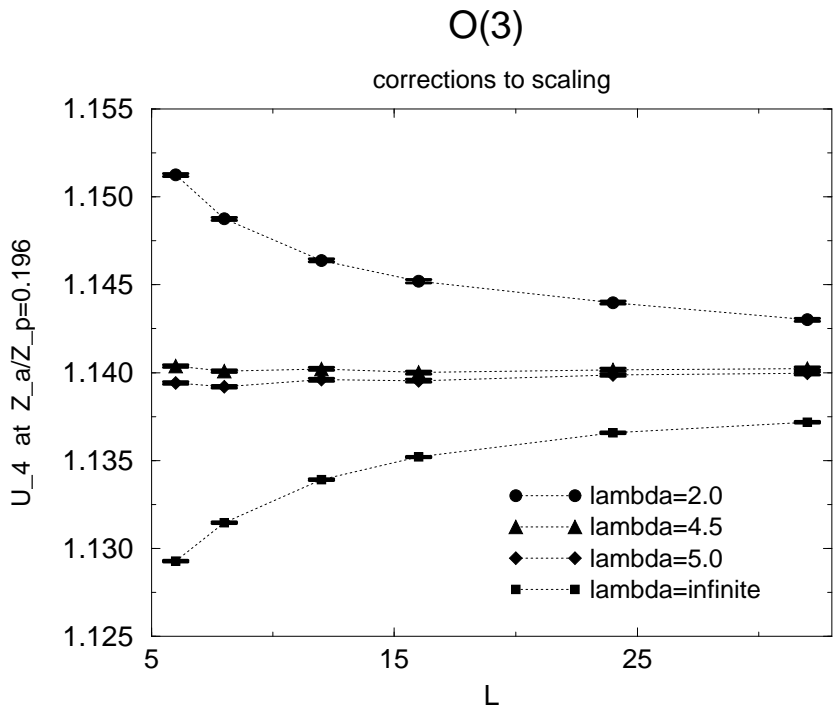}
\caption[Binder Cumulant $U$ at $Z_a/Z_p$ fixed 1]
{\label{corro3} \small
$N=3$.
The Binder-cumulant $U$ at $Z_a/Z_{p,f}=0.196$ as a function of
the lattice size $L$ for $\lambda=2.0, 4.5, 5.0$ and $\infty$.
The dotted line should only guide the eye.
}
\end{center}
\end{figure}

\begin{figure}[tb]
\begin{center}
\includegraphics[width=12cm]{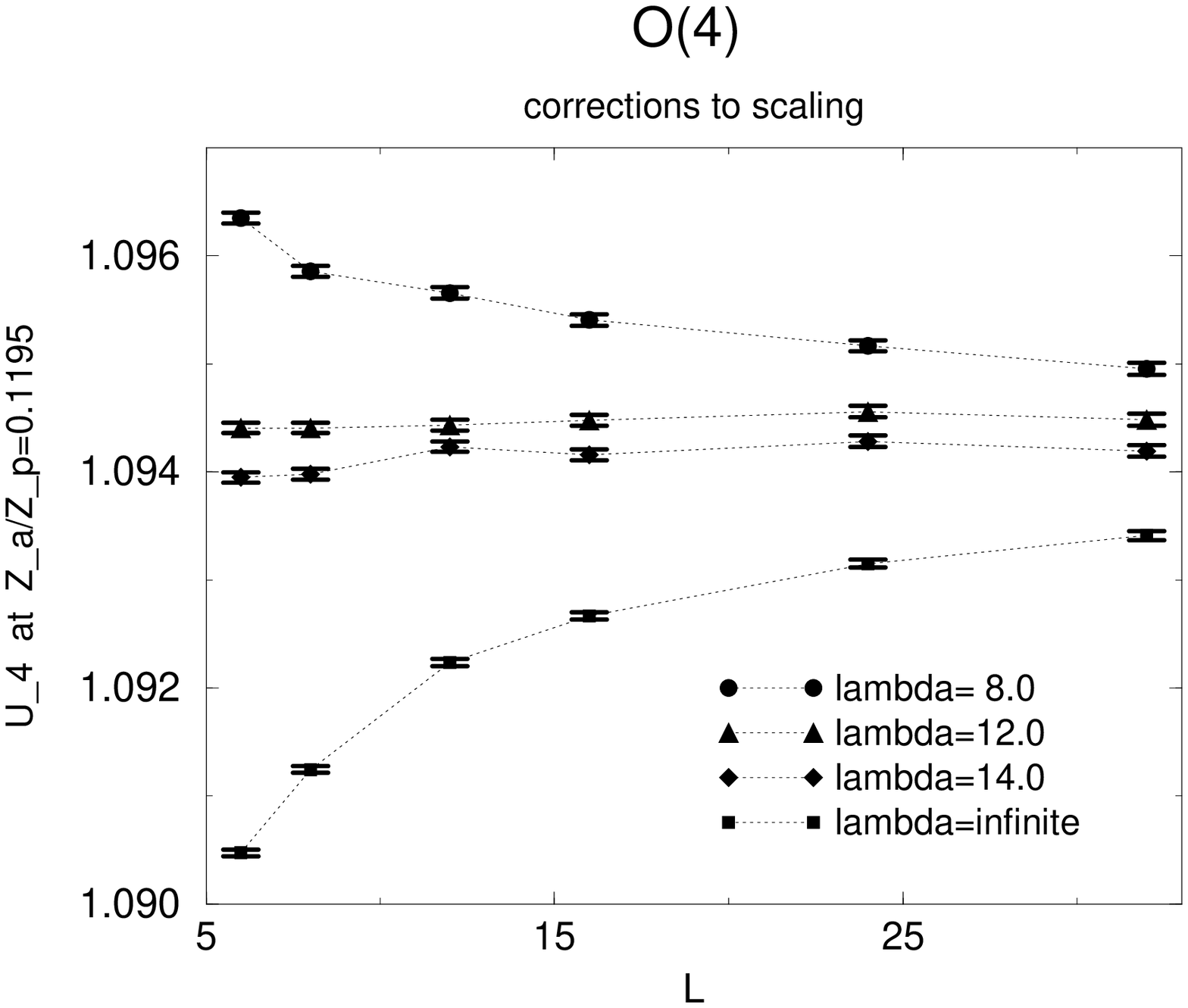}
\caption[Binder Cumulant $U$ at $Z_a/Z_p$ fixed 1]
{\label{corro4} \small
$N=4$.
The Binder-cumulant $U$ at $Z_a/Z_{p,f}=0.1195$ as a function of
the lattice size $L$ for $\lambda=8.0, 12.0, 14.0$ and $\infty$.
The dotted line should only guide the eye.
}
\end{center}
\end{figure}

Next we have analyzed $\bar{R}$ to study corrections to scaling. 
To obtain a first impression, we have plotted
our results for $\bar{R}$ with $Z_a/Z_{p,f}$ in figure \ref{corro3} for 
$N=3$ and in figure \ref{corro4} for $N=4$. In both cases the 
range $6 \le L \le 32$ is shown.

Let us discuss in detail the $N=3$ case. For the 
O(3)-symmetric non-linear $\sigma$ model
we clearly see an increase of $\bar{R}$ with increasing $L$ over the 
whole range of lattices sizes. On the other side, for $\lambda=2.0$, $\bar{R}$
is decreasing. For $\lambda=4.5$ and $\lambda=5.0$ $\bar{R}$ stays almost 
constant. This behaviour suggests that leading corrections to scaling 
vanish at $\lambda^* \approx 5$. 
The behaviour of $\bar{R}$ for $N=4$ is qualitatively the same
as for $N=3$. The figure \ref{corro4} indicates that 
$\lambda^* \approx 13$.

In the following numerical analysis of the 
data we demonstrate that the behaviour discussed above is indeed due to leading 
corrections to scaling and give an accurate estimate of $\lambda^*$ and 
its error-bar. For this purpose we fitted our data for $\bar{R}$ with the 
ansatz 
\begin{equation}
\label{corransatz}
\bar{R} \; = \; \bar{R}^* \;+\; \bar{c}(\lambda) \; L^{-\omega}
\end{equation}
with $\bar{R}^*$, $\bar{c}(\lambda)$ for each value of $\lambda$ and 
$\omega$ as free parameters. For $N=3$ and $N=4$,  we have performed such 
fits for three different sets of input data. These sets are given in table
\ref{seto34}.

\begin{table}
\caption{\sl \label{seto34}
The sets of input data for fits with eq.~(\ref{corransatz}). For $N=3$ as well
as $N=4$ we have fitted with three different sets of input data.
In column one the $\lambda$-values for $N=3$ and in  column two the 
$\lambda$-values for $N=4$ are given. In columns three, four and five the 
lattice sizes $L$ that have been included in the fits are listed.
}
\begin{center}
\begin{tabular}{|c|c|l|l|l|}
\hline
 $\lambda$, $N=3$  & $\lambda$, $N=4$ &   set 1     &  set 2    & set 3 \\
\hline
2.0 & \phantom{0}8.0 &  16, 24, 32    &  12, 16, 24    &  8, 12, 16 \\ 
4.5 & 12.0         &  16, 24, 32, 48&  12, 16, 24, 32&  8, 12, 16, 24 \\
5.0 & 14.0         &  16, 24, 32    &  12, 16, 24    &  8, 12, 16 \\
$\infty$ & $\infty$  &  16, 24, 32, 48&  12, 16, 24, 32&  8, 12, 16, 24 \\
\hline
\end{tabular}
\end{center}
\end{table}

Our results for $\bar{R}^*$ and $\omega$ for $N=3$ are given in table
\ref{omega3}.

\begin{table}
\caption{\sl \label{omega3}
Results for $\bar{R}^*$ and $\omega$ for $N=3$ from fits with the 
ansatz (\ref{corransatz}). The data-sets  that have been used for the fits 
are given in table \ref{seto34}. The results for $\bar{c}$ from the same 
fits are summarized in table \ref{barco3}.
}
\begin{center}
\begin{tabular}{|c|c|c|c|}
\hline
\rule[0mm]{0mm}{4mm}
set  & $\chi^2/$d.o.f. & $\bar{R}^*$ & $\omega$ \\
\hline
\multicolumn{4}{|l|}{$U_4$ at $Z_a/Z_{p,f}$} \\
\hline
1& 2.32 & 1.14018(9) & 0.743(23) \\
2& 1.79 & 1.14028(9) & 0.749(18) \\
3& 1.14 & 1.14021(8) & 0.799(13) \\
\hline
\multicolumn{4}{|l|}{$U_4$ at  $\xi_{2nd}/L$} \\
\hline
1& 2.16 & 1.13977(10)& 0.741(22) \\
2& 0.90 & 1.14000(11)& 0.732(17) \\
3& 4.06 & 1.14038(10)& 0.775(12) \\
\hline
\end{tabular}
\end{center}
\end{table}
The values of $\bar{c}$ from the same fits are summarized in table
\ref{barco3}.
\begin{table}
\caption{\sl \label{barco3}
Results for $\bar{c}$ for $N=3$ from fits with the ansatz (\ref{corransatz}).
The results for $\bar{R}^*$ and $\omega$ are given in table \ref{omega3}.
}
\begin{center}
\begin{tabular}{|r|l|l|l|l|}
\hline
\multicolumn{1}{|c|}{set} &
 \multicolumn{1}{c|}{$\bar{c}(2.0)$} &
 \multicolumn{1}{c|}{$\bar{c}(4.5)$} & 
 \multicolumn{1}{c|}{$\bar{c}(5.0)$} &
 \multicolumn{1}{c|}{$\bar{c}(\infty)$} \\
\hline
\multicolumn{5}{|l|}{$U_4$ at $Z_a/Z_{p,f}$} \\
\hline
 1 & 0.0392(33) &--0.0008(10) &--0.0042(8) &--0.0390(24) \\
 2 & 0.0394(25) &--0.0011(8)  &--0.0049(7) &--0.0408(17) \\
 3 & 0.0451(17) &--0.0007(6)  &--0.0053(5) &--0.0461(11) \\
\hline
\multicolumn{5}{|l|}{$U_4$ at $\xi_{2nd}/L_f$} \\
\hline
 1 & 0.0464(37) & --0.0007(11)&--0.0056(9) &--0.0464(26) \\
 2 & 0.0436(27) & --0.0026(8) &--0.0074(7) &--0.0475(18) \\
 3 & 0.0452(18) & --0.0064(6) &--0.0115(5) &--0.0566(12) \\
\hline
\end{tabular}
\end{center}
\end{table}
First of all we notice that our fit results for $\omega$ are consistent
with the estimates from field-theoretic methods. (See table \ref{crito3}.)
This indicates that the corrections are indeed dominated by the leading 
irrelevant scaling field $u_3$.  Given the statistical error and the variation
of our result for $\omega$ with the different data sets, we can not provide 
a more accurate estimate for $\omega$ as the field-theoretic methods.

Having convinced ourself that we really see leading corrections to scaling
we extract an estimate for $\lambda^*$ from the data given in table
\ref{barco3}.
Therefore, we linearly extrapolate the 
results of $\bar{c}$ at $\lambda=4.5$ and $\lambda=5.0$. Taking the results
from the set 1 we 
arrive at the estimate $\lambda^*=4.4(7)$, where the result from 
$U_4$ at $Z_a/Z_{p,f}$ and $U_4$ at $\xi_{2nd}/L_f$ are consistent. 
The error-bar is computed from the variation of $\lambda^*$ with the 
data-sets used for the fit. In the case of $U_4$ at $Z_a/Z_{p,f}$ 
$\lambda^*$ is roughly the same for all three sets. However for
$U_4$ at $\xi_{2nd}/L_f$ there is a drift to larger results for $\lambda^*$ 
as the lattice sizes increase. Assuming a convergence proportional 
$L^{-\omega_2+\omega} \approx L^{-0.8}$ we arrive at our error-estimate.

In tables \ref{omega4} and \ref{baro4} we have summarized our results for
$\bar{R}^*$ and $\omega$ and $\bar{c}$ for $N=4$. As for $N=3$ we see that 
the values for $\omega$ are consistent with the field-theoretic results.

Clearly, the sign of $\bar{c}(\infty)$ is negative and that of $\bar{c}(8.0)$ 
is positive. Hence there exists a $\lambda^*$ with $\bar{c}(\lambda^*)=0$.
From linear interpolation of the result for $\lambda=12.0$ and $\lambda=14.0$
we arrive at $\lambda^*=12.5(4.0)$. The error-bar has been computed in the 
same way as for $N=3$.

\begin{table}
\caption{\sl \label{omega4}
Results for $\bar{R}^*$ and $\omega$ for $N=4$ from fits with the 
ansatz (\ref{corransatz}). The data-sets  that have been used for the fits 
are given in table \ref{seto34}. The results for $\bar{c}$ from the same 
fits are summarized in table \ref{baro4}.
}
\begin{center}
\begin{tabular}{|c|c|c|c|}
\hline
\rule[0mm]{0mm}{4mm}
set & $\chi^2/$d.o.f. & $\bar{R}^*$ & $\omega$ \\
\hline
\multicolumn{4}{|l|}{$U_4$ at $Z_a/Z_{p,f}$} \\
\hline
1 &
0.75 &
1.09441(6) & 0.798(52) \\
2 &
0.84 &
1.09450(7) & 0.748(41) \\
3 &
2.07 &
1.09466(6) & 0.761(27) \\
\hline
\multicolumn{4}{|l|}{$U_4$ at  $\xi_{2nd}/L$} \\
\hline
1  &
0.77 &
1.09458(7) & 0.761(50) \\
2  &
1.08  &
1.09474(8) & 0.735(37) \\
3 &
7.51 &
1.09522(8) & 0.761(25) \\
\hline
\end{tabular}
\end{center}
\end{table}

\begin{table}
\caption{\sl \label{baro4}
Results for $\bar{c}$ for $N=4$ from fits with the ansatz (\ref{corransatz}).
The results for $\bar{R}^*$ and $\omega$ are given in table \ref{omega4}.
}
\begin{center}
\begin{tabular}{|r|l|r|r|r|}
\hline
 \multicolumn{1}{|c|}{set}
 & 
 \multicolumn{1}{c|}{$\bar{c}(8.0)$} &
 \multicolumn{1}{c|}{$\bar{c}(12.0)$} & 
 \multicolumn{1}{c|}{$\bar{c}(14.0)$} &
 \multicolumn{1}{c|}{$\bar{c}(\infty)$} \\
\hline
\multicolumn{5}{|l|}{$U_4$ at $Z_a/Z_{p,f}$} \\
\hline
  1    & 0.0091(19) &  0.0010(8) & --0.0023(7) & --0.0160(22) \\
  2    & 0.0073(12) &--0.0001(6) & --0.0021(5) & --0.0145(14) \\
  3    & 0.0061(7)  &--0.0013(4) & --0.0033(3) & --0.0165(8)\phantom{0} \\
\hline
\multicolumn{5}{|l|}{$U_4$ at $\xi_{2nd}/L_f$} \\
\hline
  1    & 0.0083(18) &   0.0011(8) & --0.0035(6) & --0.0183(23)\\
  2    & 0.0065(12) & --0.0017(6) & --0.0040(5) & --0.0184(15)\\ 
  3    & 0.0030(7) & --0.0059(4) & --0.0080(3)  &  --0.0239(10)\\
\hline
\end{tabular}
\end{center}
\end{table}

We have also used a slightly different approach to compute $\omega$ from 
the available data. 
We have analyzed the difference of $\bar{R}$ at $\lambda=2.0$
and $\lambda=\infty$. Obviously, $\bar{R^*}$ is cancelled this way. In addition 
one might expect that corrections that are quadratic in the scaling field $u_3$
cancel since $|\bar{c}(2.0)|$ and $|\bar{c}(\infty)|$ are almost the same.
In addition, the analysis of numerical data for $N=1$ in ref. \cite{Ha99} and 
$N=2$ in ref. \cite{HaTo99} suggests that sub-leading corrections also cancel
to a large extent. Therefore,
we have fitted our data with the ansatz
\begin{equation}
\label{diffansatz}
 \left. \bar{R}(L,\lambda)\right |_{\lambda=2.0} -
 \left. \bar{R}(L,\lambda)\right |_{\lambda=\infty} = \Delta \bar{c} \;L^{-\omega}
 \;\;\;.
\end{equation}
Our results for $N=3$ and the corresponding results for $N=4$
are given in table \ref{o34omega}. In the case of $N=4$ we have taken the 
difference of $\bar{R}$ at $\lambda=8.0$ and $\lambda=\infty$.

First we notice that a $\chi^2/$d.o.f. $\approx 1$ is already reached for 
$L_{min}=6$. For such a small $L_{min}$, fits with ansatz (\ref{corransatz})
produce  $\chi^2/$d.o.f. $= 2.8$ for $N=3$,  $U_4$ at $Z_a/Z_p=0.196$
and $\chi^2/$d.o.f. $=13.3 $ for $N=3$, $U_4$ at $\xi_2/L=0.564$. 
This fact indicates that the above 
mentioned cancellations indeed occur.

The results for $\omega$ are consistent with those of the field theoretic methods.
Certainly our approach is very promising to give competitive results for the 
correction exponent $\omega$. However, we would like to have still larger 
statistics and a larger range of lattice sizes to give a sensible estimate 
of the systematic error caused by sub-leading corrections.

\begin{table}
\caption{\sl \label{o34omega}
Results for $\omega$ for $N=3$ and $N=4$ obtained by fitting with the 
ansatz~(\ref{diffansatz}). In the first column we give $N$ and the 
phenomenological coupling that has been used to determine $\beta_f$.
In the fit all lattice size with $L_{min}\le L \le L_{max}$ have been taken 
into account.
}
\begin{center}
\begin{tabular}{|l|c|c|c|l|}
\hline
                          & $L_{min}$ & $L_{max}$ & $\chi^2/$d.o.f. & 
\multicolumn{1}{c|}{$\omega$} \\
\hline
$N=3$, $U_4$ at $Z_a/Z_p=0.196$  & 6  & 32 & 1.67 & 0.796(7) \\
                                & 8  & 32 & 0.73 & 0.781(10) \\
\hline
$N=3$, $U_4$ at $\xi_2/L=0.564$  & 6  & 32 & 0.65 & 0.769(6) \\ 
                                & 8  & 32 & 0.65 & 0.766(9) \\
\hline
\hline
$N=4$, $U_4$ at $Z_a/Z_p=0.1195$ & 6 & 32 & 0.54 & 0.780(15) \\
                                & 8 & 32 & 0.40 & 0.765(22) \\
\hline
$N=4$, $U_4$ at $\xi_2/L=0.547$  & 6 & 32 & 0.40 & 0.774(14) \\
                                & 8 & 32 & 0.38 & 0.764(20) \\
\hline
\end{tabular}
\end{center}
\end{table}

\subsection{Critical exponents}
\label{exponents}
We have computed the critical exponents $\nu$ and $\eta$ using well
established finite size scaling methods.  
Below we shall only discuss in detail the results for $N=3$. The analysis for $N=4$
has been performed analogously.

The exponent $\nu$ is computed from 
the slope of a phenomenological coupling $R$ at $\beta_f$ 
(see eq.~(\ref{betafeq})):
\begin{equation}
\label{nuansatz}
 \left . \frac{\partial R}{\partial \beta} \right |_{\beta_f} \; = 
 \; a \; L^{1/\nu} \;\;\;.
\end{equation}
As it was pointed out in \cite{Ballesteros},
replacing $\beta_c$ by $\beta_f$ simplifies the error-analysis, since the 
error in $\beta_c$ needs not to be propagated.

In our study we have considered three different choices of $R$. Hence in 
eq.~(\ref{nuansatz}) we could in principle consider 9 different combinations.
Below we shall restrict ourself to six choices: $\beta_f$ is fixed either 
by $Z_a/Z_{p,f}$ or $\xi_{2nd}/L_f$. We consider the slope of all three 
phenomenological couplings $R$.

First we like to study how much the result of $\nu$ from fits with
eq.~(\ref{nuansatz}) depends on leading corrections to scaling.
For this purpose we have fitted for $L_{min}=16$ and $L_{max}=32$ our data
for all available values of $\lambda$ for all six combinations of $R$.
The results of these fits are summarized in table \ref{nucorrections}.

The variation of the results with $\lambda$ are rather small. The largest 
variation
we find for the combination C1, where we get $\nu=0.7164(23)$ for $\lambda=2.0$
and $\nu=0.7076(12)$ for $\lambda=\infty$. Hence, we expect that for 
$\lambda=4.5$ the effect of 
corrections to scaling  should be smaller than $0.0001$.

\begin{table}
\caption{\sl \label{nucorrections}
 Results for $\nu$ from fits with the ansatz (\ref{nuansatz}). 
 The range of lattice sizes is always $L_{min}=16$ and $L_{max}=32$. 
 We have used six different combinations of $R$. These are given by
  C1: The slope of $U_4$ at $Z_a/Z_{p,f}$;
  C2: The slope of $Z_a/Z_p$ at $Z_a/Z_{p,f}$; 
  C3: The slope of $\xi_{2nd}/L$ at  $Z_a/Z_{p,f}$; 
  C4: The slope of $U_4$ at $\xi_{2nd}/L_f$;
  C5: The slope of $Z_a/Z_p$ at $\xi_{2nd}/L_f$;
  C6: The slope of $\xi_{2nd}/L$ at $\xi_{2nd}/L_f$.
 }
\begin{center}
\begin{tabular}{|c|l|l|l|l|l|l|}
\hline
$\lambda$ &  \multicolumn{1}{c|}{C1}       & 
             \multicolumn{1}{c|}{C2}       & 
	     \multicolumn{1}{c|}{C3}       & 
	     \multicolumn{1}{c|}{C4}       &  
	     \multicolumn{1}{c|}{C5}       &  
	     \multicolumn{1}{c|}{C6}  \\
\hline
  2.0    & 0.7164(23) & 0.7088(10)& 0.7136(13)& 0.7181(24)& 0.7106(12)& 0.7135(12) \\
  4.5    & 0.7071(21) & 0.7083(9) & 0.7115(12)& 0.7074(21)& 0.7086(10)& 0.7115(12) \\
  5.0    & 0.7078(21) & 0.7085(9) & 0.7114(11)& 0.7071(21)& 0.7077(10)& 0.7114(11) \\
$\infty$ & 0.7076(12) & 0.7127(5) & 0.7142(6) & 0.7056(12)& 0.7136(13)& 0.7141(6) \\
\hline
\end{tabular}
\end{center}
\end{table}

Next, let us discuss in more detail the results for $\lambda=4.5$  which is 
closest to our estimate of $\lambda^*$. Results from fits with ansatz 
(\ref{nuansatz}) are summarized in table \ref{nufinal}. We see that the results
approach each other as $L_{min}$ is increased. In the case of the slope of $U_4$,
the results for $\nu$ stay almost constant as $L_{min}$ is varied. For the 
slope of $Z_a/Z_p$ we see a slight increase of the estimate of $\nu$. On the 
other hand for the slope of $\xi_{2nd}/L $ we see a decrease. Assuming that this
behaviour is caused by the sub-leading corrections,
we conclude that $\nu=0.7120$
from $\xi_{2nd}/L $ is an upper bound. Given the larger stability 
of the estimate
from  $Z_a/Z_p$ we take as our final estimate $\nu=0.710(2)$.

\begin{table}
\caption{\sl \label{nufinal}
Results for $\nu$ from fits with ansatz~(\ref{nuansatz}) 
for $N=3$ at $\lambda=4.5$; always $L_{max}=48$.
The combinations C1,...,C6 are explained in the caption of table 
\ref{nucorrections}.
}
\begin{center}
\begin{tabular}{|c|l|l|l|l|l|l|}
\hline
 $L_{min}$ &  \multicolumn{1}{c|}{C1}       & 
  \multicolumn{1}{c|}{C2}       &  
  \multicolumn{1}{c|}{C3}       &  
  \multicolumn{1}{c|}{C4}       &   
  \multicolumn{1}{c|}{C5}       &   
 \multicolumn{1}{c|}{C6}  \\
\hline
 8   & 0.7113(7)  & 0.7087(3) & 0.7150(4) & 0.7102(7) & 0.7075(3) & 0.7150(4)\\
12   & 0.7111(10) & 0.7100(4) & 0.7132(5) & 0.7107(10)& 0.7096(5) & 0.7132(5)\\
16   & 0.7101(13) & 0.7099(6) & 0.7120(7) & 0.7102(14)& 0.7100(7) & 0.7120(7)\\
\hline
\end{tabular}
\end{center}
\end{table}

Next we computed the exponent $\eta$ from the finite size behaviour of the 
magnetic susceptibility
\begin{equation}
\label{etaansatz}
\left . \chi  \right |_{\beta_f} \; = 
\; c \; L^{2-\eta} \;\;\;.
\end{equation}
We restrict the discussion to $Z_a/Z_{p,f}$ since fixing $\beta_f$ by 
$\xi_{2nd}/L_f$ gives very similar numbers.

For the estimate of $\eta$ we see a much stronger dependence on $\lambda$ 
than for $\nu$. Fitting with ansatz (\ref{etaansatz}) and $L_{min}=16$ , 
$L_{max}=32$ we obtain $\eta=0.03982(33)$, $0.03614(30)$, $0.03541(31)$ and 
$0.03269(18)$ for $\lambda=2.0$, $4.5$, $5.0$ and $\infty$, respectively.

Fitting the data at $\lambda=4.5$ with $L_{max}=48$ yields $\eta=0.03581(14)$, 
$0.03668(19)$ and $0.03736(32)$ for $L_{min}=12$, $16$ and $24$, respectively.
For $L_{min}=12$ we get $\chi^2/$d.o.f.$ = 18.5$. The strong dependence of 
the result on $L_{min}$ and the large  $\chi^2/$d.o.f. at $L_{min}=12$ indicates 
that there are sizeable sub-leading corrections.

Fitting with an ansatz that includes an analytic background term
\begin{equation}
 \left . \chi  \right |_{\beta_f} \; =
 \; c \; L^{2-\eta} \;+\; b \;\;
\end{equation}
yields  $\chi^2/$d.o.f.$ = 0.38$ already for $L_{min}=8$. The results are
$\eta=0.0384(2)$, $0.0386(4)$ and $0.0381(6)$ for $L_{min}=8$, $12$ and $16$, 
respectively.
To see the effect of leading corrections on this fit we 
have in addition fitted the data for 
$\lambda=5.0$ with $L_{min}=8$ and $L_{max}=32$. 
We get $\eta=0.0378(4)$.

As our final estimate we quote $\eta=0.0380(10)$. The error-bar takes into account
statistical errors as well a systematic errors due to leading and sub-leading 
corrections. These errors are estimated from the spread of the results
of the various fits discussed above.

With a similar analysis we arrive at $\nu=0.749(2)$ and $\eta=0.0365(10)$ 
for the $O(4)$ universality class.

\section{Comparison with results from the literature}
Here we like to compare our results for the critical exponents $\nu$ and 
$\eta$ with previous Monte Carlo studies of the $O(3)$- and $O(4)$-invariant
non-linear $\sigma$ models. In addition we give selected results 
from high temperature series, perturbation theory in three dimensions, and 
the $\epsilon$-expansion. The results are summarized in tables \ref{crito3}
and  \ref{crito4} for $N=3$ and $N=4$, respectively. For more references  
on field theoretic methods and other methods, not discussed here, see e.g.
ref. \cite{GUZI}.

All Monte Carlo studies listed in the tables \ref{crito3} and  \ref{crito4}
use a simple cubic lattice. In addition, 
in ref. \cite{ChFeLa} the body centred cubic lattice is studied. (In table
\ref{crito3} we only give the sc results.)
In all studies the lattice sizes are smaller or equal $L=48$, except for 
ref. \cite{Ballesteros}, where in addition $L=64$ is simulated.
The results of the MC studies cited above are extracted from ans\"atze like 
eqs.~(\ref{nuansatz},\ref{etaansatz}). 
(Mostly $\beta_c$ is used instead of $\beta_f$.)
We see that almost all Monte Carlo
results for $\nu$ are consistent with ours. On the other hand, the results
for $\eta$ are systematically too small, except for ref. \cite{Ballesteros}.

This behaviour can be well understood with our results of section 
\ref{exponents}. The estimates for $\eta$ from the ansatz (\ref{etaansatz})
are clearly effected by corrections to scaling. Our results from the 
$O(3)$- and $O(4)$-invariant non-linear  $\sigma$ models for $\eta$ are
systematically lower than our final results from the $\phi^4$ models at 
$\lambda^*$.  On the other hand, the results for $\nu$ given in table
\ref{nucorrections} show only little variation with $\lambda$; i.e. little
dependence on leading corrections to scaling.

The authors of ref. \cite{Ballesteros}, who for the first time tried 
to take into account leading corrections to scaling in the analysis of their
data, arrive at rather 
similar conclusions how leading corrections to scaling affect the 
estimates of $\eta$ and $\nu$. They extrapolated their results for
$\eta$ assuming $L^{-\omega}$ corrections. 

However from our analysis of section \ref{exponents} we know that 
the estimates of $\eta$ obtained from lattices with $L\le48$ are 
also strongly affected by sub-leading corrections with $\omega_2\approx 2$.

Hence, extrapolating only in  $L^{-\omega}$ leads to a wrong 
amplitude for the $L^{-\omega}$ corrections. As  result, the 
final estimate of ref. \cite{Ballesteros} for $\eta$ is too large compared
with our result.

There exists a large number of refs. on the $\epsilon$-expansion and 
the perturbative expansion in three dimensions in the literature. As an 
example, we have chosen the result of a recent analysis by 
Guida and Zinn-Justin \cite{GUZI}. We notice that these results are 
consistent with ours for $\eta$ and $\nu$  within the quoted errors.
Also note that the error-bars of our estimates for $\nu$ and $\eta$  
are smaller than those of the field-theoretic estimates.

There exists also a number of publications on the analysis of high 
temperature series.
In the tables we give the results of a recent 
analysis \cite{BuCo2} 
using inhomogeneous differential approximants. The coefficients of the high
temperature series of $\chi$ and $\mu_2$ are computed up to $\beta^{21}$.
In our tables, we only give the results from the unbiased analysis of 
the simple cubic lattice
series. In addition the authors analyze the body centred cubic lattice. 
The authors also give results obtained from a so called $\theta$-biased analysis, 
where they make use of the numerical results for $\theta=\omega \nu$ obtained 
from field-theoretic methods. It is interesting to notice that these baised
results (not given in our tables) tend to be less consistent with our results 
than the un-biased results which we quote in tables \ref{crito3} and \ref{crito4}.

\begin{table}
\caption{\sl \label{crito3}
Results for the critical exponents of the $O(3)$ universality class 
from various methods. The numbers for $\eta$ that are marked by a $*$ 
are computed from $\gamma/\nu=2-\eta$. Details are discussed in the text.
}
\begin{center}
\begin{tabular}{|c|c|l|l|l|}
\hline
 method              &ref.       & \multicolumn{1}{c|}{$\nu$}    & 
				   \multicolumn{1}{c|}{$\eta$} & 
				   \multicolumn{1}{c|}{$\omega$} \\
\hline
IMC                  &present work      &0.710(2)  & 0.0380(10)   & 0.773    \\  
 MC                  &\cite{Ballesteros}&0.7128(14)& 0.0413(15)(1)& 0.78(2) \\
 MC                  &\cite{BrCi}      & 0.642(2)  & 0.020(1)     &         \\ 
 MC                  &\cite{JaHo93}      &0.704(6) & 0.027(2)     &         \\
 MC                  &\cite{JaHo92}      &0.704(6) & 0.028(2)     &         \\
 MC                  &\cite{PeFeLa}       &0.706(9) & 0.031(7)     &        \\
 MC                  &\cite{ChFeLa}     & 0.7036(23)& 0.0250(35)  &         \\
 HT                  &\cite{BuCo2}       &0.715(3)  & $0.036(10)^*$ &       \\
 $d=3$ PT            & \cite{GUZI}      &0.7073(35)& 0.0355(25)   & 0.782(13) \\
$\epsilon$-expansion & \cite{GUZI}      &0.7045(55)& 0.0375(45)   & 0.794(18) \\
\hline
\end{tabular}
\end{center}
\end{table}

\begin{table}
\caption{\sl \label{crito4}
Results for the critical exponents of the $O(4)$ universality class 
from various methods. The numbers for $\eta$ that are marked by a $*$ 
are computed from $\gamma/\nu=2-\eta$. Details are discussed in the text.
}
\begin{center}
\begin{tabular}{|c|c|l|l|l|}
\hline
 method              &ref.              & \multicolumn{1}{c|}{$\nu$}    & 
					  \multicolumn{1}{c|}{$\eta$} & 
					  \multicolumn{1}{c|}{$\omega$} \\
\hline
 IMC                 &present work     & 0.749(2)  & 0.0365(10) & 0.765   \\
 MC                  &\cite{Ballesteros}& 0.7525(10)& 0.0384(12) & 1.8(2)  \\
 MC                  &\cite{KaKa}      & 0.7479(90) & 0.0254(38) &        \\
  HT                  &\cite{BuCo2}    & 0.750(3) & $0.035(9)^*$ &      \\       
 $d=3$ PT            &\cite{GUZI}       & 0.741(6)  & 0.0350(45) & 0.774(20) \\
$\epsilon$-expansion &\cite{GUZI}       & 0.737(8)  & 0.0360(40) & 0.795(30) \\
\hline
\end{tabular}
\end{center}
\end{table}

In tables  \ref{crito3} and \ref{crito4}, 
we give for $\omega$ the average of the result
from $Z_a/Z_{p,f}$ and $\xi_{2nd}/L_f$  with $L_{min}=8$ taken from table 
\ref{o34omega}. We make no attempt to estimate the systematic errors  
due to sub-leading corrections.
Certainly these errors are larger than that quoted for the 
field-theoretic estimates of ref. \cite{GUZI}. It is however interesting to note 
that our results are consistent with those of ref. \cite{GUZI}.

Also the Monte Carlo result of ref. \cite{Ballesteros} for $N=3$ is consistent with 
ours. However we cannot confirm their surprising result for $N=4$.

\section{Conclusions}
In this study we have demonstrated that the programme of refs. 
\cite{spain,HaPiVi} to eliminate leading corrections to scaling in the three
dimensional $\phi^4$ model can be extended to $N=3$ and $N=4$.  In particular,
we have found $\lambda^* =4.4(7)$ for $N=3$ and $\lambda^* =12.5(4.0)$ 
for $N=4$.  Based on this result we have computed the critical exponents 
$\nu$ and $\eta$ from finite size scaling. In particular in the case of $\eta$, 
the error-bar 
could be reduced considerably compared with previous Monte Carlo simulations 
or field theoretic methods and the analysis of high temperature series.

Since the CPU-time that was used for the present study is still 
moderate, further progress can be made by just enlarging the statistics and 
simulating larger lattices.

Also, our results for $\lambda^*$ can be used as input for the analysis 
of high temperature series analogous to refs. \cite{pisa,pisaXY}. 

The principle question raised in ref. \cite{pisa}, 
whether the programme to eliminate leading corrections is restricted to 
$N< N_c$, where $N_c$ is finite, remains open.

\section{Acknowledgements}
I like to thank M. Campostrini, A. Pelissetto, P. Rossi and E. Vicari 
for discussions and comments on the manuscript. I also like to thank 
M. M\"uller-Preu{\ss}ker for giving me access to the PC-pool of his 
group.


\begin{thebibliography}{99}

\bibitem{WiKo}
K. G. Wilson and J. Kogut, Phys. Rep. C 12 (1974) 75.   

\bibitem{cardy}
J.Cardy, {\sl
Scaling and Renormalization in Statistical Physics},
(Cambridge University Press,
1996).

\bibitem{Domb}
C. Domb, {\sl The Critical Point, A historical introduction to the modern
theory of critical phenomena}, (Taylor$\&$Francis, 
1996).

\bibitem{ChFiNi}
J. H. Chen, M. E. Fisher and B. G. Nickel, Phys. Rev. Lett. 48 (1982) 630; \\
M. E. Fisher and J. H. Chen, J. Physique (Paris) 46 (1985) 1645.

\bibitem{NiRe}
B. G. Nickel and J. J. Rehr, J. Stat. Phys. 61 (1990) 1.

\bibitem{spain}
H. G. Ballesteros, L. A. Fernandez, V. Martin-Mayor and A. Munoz-Sudupe, \\
hep-lat/9805022, Phys. Lett. B 441 (1998) 330.

\bibitem{HaPiVi}
M. Hasenbusch, K. Pinn and S. Vinti, cond-mat/9804186, unpublished\\
M. Hasenbusch, K. Pinn and S. Vinti, 
hep-lat/9806012, Phys. Rev. B 59 (1999) 11471.

\bibitem{Ha99}
M. Hasenbusch, hep-lat/9902026, J. Phys. A. 32 (1999) 4851.

\bibitem{HaTo99}
M. Hasenbusch and T. T\"or\"ok, cond-mat/9904408,
J. Phys. A 32 (1999) 6361.

\bibitem{XYneu}
M. Campostrini, M. Hasenbusch, A. Pelissetto, P. Rossi and E. Vicari,
cond-mat/0010360.

\bibitem{pisa}
M. Campostrini, A. Pelissetto, P. Rossi and E. Vicari,
cond-mat/9905078, Phys. Rev. E 60 (1999) 3526.

\bibitem{pisaXY}
M. Campostrini, A. Pelissetto, P. Rossi and E. Vicari,
cond-mat/9905395,
Phys. Rev. B 61 (2000) 5905, \\
M. Campostrini, A. Pelissetto, P. Rossi and E. Vicari,
cond-mat/0001440.

\bibitem{milano}
P. Butera and M. Comi, hep-lat/9805025, Phys. Rev. B 58 (1998) 11552.

\bibitem{Nightingale}
M. P. Nightingle, Physica A 83 (1976) 561.   

\bibitem{binder}
K. Binder, Z. Phys. B 43 (1981) 119;
K. Binder, Phys. Rev. Lett. 47 (1981) 693. 

\bibitem{GUZI}
R. Guida and J. Zinn-Justin,
cond-mat/9803240, J. Phys. A 31 (1998) 8103.

\bibitem{NR-84}
K. E. Newman and E. K. Riedel, Phys. Rev. B 30 (1984) 6615.

\bibitem{CPRV-98}
M. Campostrini, A. Pelissetto, P. Rossi, and E. Vicari,
Phys. Rev. E  57 (1998) 184.

\bibitem{BrTa}
R. C. Brower and P. Tamayo,
Phys. Rev. Lett. 62 (1989) 1087.

\bibitem{SW}
R. H. Swendsen and J.-S. Wang,
Phys. Rev. Lett. 58 (1987) 86.

\bibitem{Wolff}
U. Wolff, Phys. Rev. Lett. 62 (1989) 361.

\bibitem{Ha93}
M. Hasenbusch, hep-lat/9209016, J. Phys. I France 3 (1993) 753.

\bibitem{GH-94}
A. P. Gottlob and M. Hasenbusch,
cond-mat/9406092, J. Stat. Phys. 77 (1994) 919. 



\bibitem{Ballesteros}
H. G. Ballesteros, L. A. Fernandez, V. Martin-Mayor, A. Munoz Sudupe, 
cond-mat/9606203,
Phys. Lett. B387 (1996) 125.

\bibitem{BrCi}
R.G. Brown and M. Ciftan, Phys. Rev. Lett. 76 (1996) 1352.

\bibitem{JaHo93}
C. Holm and  W. Janke, hep-lat/9301002,
Phys. Rev. B48 (1993) 936.

\bibitem{JaHo92}
C. Holm and  W. Janke, hep-lat/9209017,
Phys. Lett. A173 (1993) 8.

\bibitem{PeFeLa}
P. Peczak, A.M. Ferrenberg, and D.P. Landau, Phys. Rev. B 43 (1991) 6087.

\bibitem{ChFeLa}
K. Chen, A.M. Ferrenberg, and D.P. Landau, Phys. Rev. B 48 (1993) 3249.

\bibitem{BuCo2} P. Butera and M. Comi, Phys.Rev. B56 (1997) 8212.
\bibitem{KaKa}
K. Kanaya and S. Kaya, Phys. Rev. D 51 (1995) 2404.


\end{thebibliography}
\end{document}